\documentclass{elsart}

 \usepackage{graphicx}

\usepackage{amssymb}

\begin{document}

\begin{frontmatter}



\title{Millisecond newly born pulsars as efficient accelerators of electrons}


\author{Zaza Osmanov$^1$,} \author{Swadesh Mahajan$^2$,} \author{George Machabeli$^3$,} \author{Nino Chkheidze$^3$}

\address{1. School of Physics, Free University of Tbilisi, 0183-Tbilisi, Georgia}

\address{2. Institute for Fusion Studies, The University of Texas at
Austin, Austin, Texas 78712}

\address{3. Centre for Theoretical Astrophysics, ITP, Ilia State
University, 0162-Tbilisi, Georgia}

\begin{abstract}
The newly born millisecond pulsars are investigated as possible
energy sources for creating ultra-high energy electrons. The
transfer of energy from the star rotation to high energy electrons
takes place through the Landau damping of centrifugally driven (via
a two stream instability) electrostatic Langmuir waves. Generated in
the bulk magnetosphere plasma, such waves grow to high amplitudes,
and then damp, very effectively, on relativistic electrons driving
them to even higher energies. We show that the rate of transfer of
energy is so efficient that no energy losses might affect the
mechanism of particle acceleration;  the electrons might achieve
energies of the order of $10^{18}$eV for parameters characteristic
of a young star.
\end{abstract}

\begin{keyword}
acceleration of particles.


\end{keyword}

\end{frontmatter}


{\bf Introduction}

One of the most fascinating stories in astro-particle physics is the
discovery of high energy cosmic ray (CR) electrons. Two important
constituents of this story are: the H.E.S.S (High Energy
Stereoscopic System) team's \cite{hesstev} claim of the detection of
TeV CR electrons of local origin (within $\sim 1$kpc), and the later
confirmation of the energy range (.007-1Tev) by the  Fermi-LAT
collaboration that  analyzed  the CR electron/positron data samples
collected from $4$ August $2008$ to $4$ August $2009$.
\cite{fermitev}.

How do particles acquire such enormous amounts of energy? In
astro-particle physics, it is conventional to invoke a Fermi-type
process: the stochastic acceleration of particles interacting with
strong shocks. Though quite effective, such a mechanism requires a
pre acceleration of particles to already high energies \cite{bell}.
Accelerating light particles like electrons to high energies is even
more problematic because the strong synchrotron radiation losses
will limit the maximum attainable energy \cite{blandford}.

One can imagine electrons being boosted to PeV or higher energies if
strong energy losses could be, somehow, prevented during the
accelerating process. One such mechanism in which the rotation
energy of  a neutron star was utilized to strongly accelerate
particles ( via Langmuir waves generated in magnetospheres of
pulsars), was recently developed and investigated. It was shown the
acceleration process proceeds with almost negligible energy losses.
\cite{screp}. The principle steps in the above mentioned process
are:

1)The rotational slow down of a pulsar provides the energy to excite
Langmuir waves in the bulk electron-positron plasma in the star
atmosphere. All pulsars are characterized by a decreasing rate of
spinning, $\dot{P}\equiv dP/dt>0$, where $P$ is the period of
rotation.The period is measured by $\dot{W} =
I\Omega|\dot{\Omega}|$, were $I=2MR_{\star}^2/5$ is the moment of
inertia of the neutron star, $M\sim 1.5\times M_{\odot}$ and
$M_{\odot}\approx 2\times 10^{33}$g are the pulsar's mass and the
solar mass respectively, $R_{\star}\approx 10^6$cm is the pulsar's
radius, $\Omega\equiv 2\pi/P$ is its angular velocity and
$\dot{\Omega}\equiv d\Omega/dt=-2\pi\dot{P}/P^2$.

2) The excited Langmuir waves , then, efficiently damp,
preferentially on the much faster but local beam electrons to
accelerate them to even larger energies

We refer the reader to (\cite{screp}, and references therein) for
the relevant pulsar phenomenology, and details of this acceleration
mechanism; henceforth, the latter twould be termed
Langmuir-Landau-Centrifugal Drive (LLCD). LLCD is based on the
centrifugal acceleration that has been successfully applied to some
astrophysical settings \cite{bogoval2,osm7}. It is also believed
that young pulsars might be sources of ultra high energy cosmic rays
\cite{youngPL}. For the paper to be self-sufficient, we will give a
synopsis of the theoretical underpinnings of LLCD before applying it
to the newly born millisecond pulsars; the results are spectacular;
LLCD can drive electron energies all the way to $10^{18}$eV.

In this paper, we do not  investigate the aftermath of the
acceleration era- for instance, the observational patterns caused by
the presence of ultra-high energy particles in the magnetospheres of
millisecond pulsars. One does expect that electrons with such
enormous energies will leave their  "fingerprints" on the emission
patterns of the corresponding pulsars. The scope of this work could
be extended to examine the effects of the electromagnetic radiation
that must come from the ultra relativistic electrons; we intend to
deal with these problems in near future. Another class of
potentially interesting problems ( in the context of the present
work) are related to the recently discovered  events of PeV
neutrinos by the Ice Cube experiment. Indeed, electrons and
positrons with energies much higher than PeV, might produce PeV
neutrinos via several channels: $e^{-}+e^{+}\rightarrow
\nu+\overline{\nu}$; $e^{-}+p\rightarrow n+\nu$.


{\bf Results}

The Langmuir-Landau-Centrifugal Drive, derived within the framework
of a relatively simple but nontrivial theoretical model, is  shown
to work highly efficiently in the young millisecond pulsars. LLCD,
through a two step process, converts the pulsar spin-down energy
into the kinetic energy of electrons.  In the first step, the
rotation imparted centrifugal particle energy is converted, via a
parametrically driven "two stream instability", to electrostatic
Langmuir waves in the bulk electron - positron plasma residing in
the pulsar magnetosphere. Landau damping of these centrifugally
excited electrostatic waves on the high energy primary beam
electrons (beam) transiting through the same region, constitutes the
second step.

The linear growth rates of the rotation driven Langmuir instability,
calculated for the newly born pulsars, are faster than the typical
rotation timescales of particles. The Langmuir instability, thus, is
efficient and can rapidly convert the star slow down energy into the
electric field energy

Landau damping of the unstable Langmuir waves on primary beam
electrons that converts the electric field energy into particle
energy is also shown to be equally rapid. The combination creates a
very efficient "machine" that generates ultra high energy (up to
EeV) cosmic ray electrons.

We will show that the LLCD mechanism, when applied to newly born
fast spinning stars could, indeed, create electron energies that
match the highest observed in the CR electrons; we believe that this
result has much significance for  high energy astro-particle
physics.



{\bf Discussion}

It is generally believed that to attain the possible maximum
possible energy, the gyroradius of the particle should be contained
in the acceleration zone. Then, the combination of a strong rotation
and a strong magnetic field (of a pulsar) leads to an enormous
induced electric field on the light cylinder surface $\Omega
rB/c\sim 10^{10}$statvolts/cm. This field, might potentially
accelerate electrons up to energies of the order of $E_{max}\approx
1.3\times 10^{19}\times B_{lc,10}\times P_{-3}$eV \cite{youngPL},
where $B_{lc,10}\equiv B_{lc}/(10^{10}Gauss)$, $P_{-3}\equiv
P/(10^{-3}s)$. What one needs , then,  is an efficient agency  in
the rotating magnetosphere that will convert the rotational energy
into the acceleration of electrons. We believe that the LLCD does
exactly that:  the Landau damping of centrifugally induced Langmuir
waves guarantees  a high efficient channel for acceleration

For these waves to effectively transfer their energy to particles,
the waves phase velocity must be close to the particle speed, which,
in this case, approaches the speed of light. Further, in the
vicinity of $\upsilon_{ph}$, there should be more particles a little
slower than the wave than particles which are a little faster (in
the opposite situation, the wave will feed off the particles). For
the given problem it is always possible to situate $\upsilon_{ph}$
in the desired part  of the  primary beam spectrum. Since the
distribution function decreases with the Lorentz factor, the number
of electrons with $\upsilon_b<\upsilon_{ph}$ exceeds that of the
electrons with $\upsilon_b>\upsilon_{ph}$, where $\upsilon_b$
denotes the electron speed. Thus the optimum conditions for
effective Landau damping and, therefore, of net energy transfer from
the star spin down to electrons, will  pertain.

For relativistic plasmas it has been shown that the Landau damping
rate is given by \cite{vkm}
\begin{equation}
\label{dampr}  \Gamma_{LD}=
\frac{n_{_{GJ}}\gamma_b\omega_b}{n_p\gamma_p^{5/2}},
\end{equation}
where $n_p$ is the plasma density and $n_{_{GJ}}\equiv
B_{lc}/(Pce)\approx 1.8\times 10^{12}$cm$^{-3}$ is the
Goldreich-Julian number density. For the primary beam,  $\gamma_b$
and $\omega_b=\sqrt{4\pi e^2 n_{_{GJ}}/m}$ are, respectively, the
Lorentz factor and the plasma frequency. Landau damping is the
second step of the acceleration mechanism LLCD.

We now demonstrate the efficiency of LLCD  for young millisecond
pulsars for which the typical period and slowdown rates are: $P\sim
10^{-3}$s and $\dot{P} \sim 10^{-12}$ss$^{-1}$. It is worth noting
that, at this stage such pulsars are only, theoretically, predicted
\cite{carroll}; this type of pulsars have not yet been observed.

The efficiency of the LLCD must depend upon the proper coordination
of the damping rates on the beam electrons and the growth rates of
Langmuir  waves in the bulk plasma. In \cite{screp}, the
magnetospheric plasma was modeled as consisting of two streams with
the Lorentz factors $\gamma_{1,2}$ ($\gamma_1<\gamma_2$). It was
shown that the growth rate of the instability is well approximated
by the expression
 \begin{equation}
 \label{grow}
 \Gamma= \frac{\sqrt3}{2}\left (\frac{\omega_1 {\omega_2}^2}{2}\right)^{\frac{1}{3}}
 {J_{\mu_{res}}(b)}^{\frac{2}{3}},
\end{equation}
where $\omega_{1,2}\equiv\sqrt{8\pi e^2n_{1,2}/m\gamma_{1,2}^3}$ are
the plasma frequencies of the two species, $n_{1,2}$ are the
corresponding number densities, $J_{\mu}(x)$ is the Bessel function
of the first kind, $b = (2ck/\Omega)\sin\phi_{-}$, $2\phi_{-} =
\phi_p -\phi_e$ and $\phi_p$ and $\phi_e$ are, respectively, the
positron's and electron's initial phases.

It stands to reason that if the instability growth rate was much
greater than the Landau damping rate, there will be little effective
energy transfer from waves to the particles. In the diagonally
opposite limit  with the damping rates far in excess of the growth
rates, the waves will not grow much, again resulting in very little
transfer from the star rotation to the waves. The most optimum
scenario for an overall efficient energy pumping/transfer system,
therefore, is realized when the instability growth and Landau
damping rates are large (with respect to the kinematic rate,
$\Omega$) and comparable, $\Gamma\sim\Gamma_{LD}$. For the
two-stream instability, the aforementioned condition can be
satisfied for particular combinations of $\gamma_1$ and $\gamma_2$;
for example, for $\gamma_1\approx 1.8\times 10^5$, $\gamma_2\approx
8\times 10^5$.

The total energy gained by the beam particles has been estimated to
be \cite{screp}
\begin{equation}
\label{en}  \epsilon\approx \frac{n_pF_{reac}\delta r}{n_{_{GJ}}},
\end{equation}
where $\delta r\sim c/\Gamma$, $F_{reac}\approx2mc\Omega\xi
(r)^{-3}$, $\xi (r) = \left(1-\Omega^2r^2/c^2\right)^{1/2}$
\cite{grg}. From Eq. (\ref{en}) it is clear that two streams with
$\gamma_p\equiv\gamma_1\approx 1.8\times 10^5$, $\gamma_2\approx
8\times 10^5$, which in turn guarantee the condition,
$\Gamma\sim\Gamma_{LD}$, lead to the acceleration of electrons up to
energies of the order of $10^{18}$eV. This estimate invokes
equipartition of energy in the three relevant sp constituents:
$n_1\gamma_1\approx n_2\gamma_2\approx n_{_{GJ}}\gamma_b$ with
$\gamma_b\approx 7.5\times 10^7$, where $n_{_{GJ}}\equiv
B/(Pce)\approx 1.8\times 10^{12}$cm$^{-3}$ is the Goldreich-Julian
number density \cite{gj} in regions in the vicinity of the light
cylinder.

Because the highly relativistic particles will, inevitably, loose
energy due to radiation, one must investigate how radiation will
affect overall energy transfer. Could radiation, for example, put a
stringent limit on the maximum energy acquired?

For highly relativistic electrons and photons with
$\epsilon\epsilon_{ph}/(m^2c^4)>1$ ($\epsilon_{ph}$ is photon
energy), the inverse Compton mechanism  operates in  the
Klein-Nishina (KN) regime \cite{Lightman}. Energy emitted per
particle per second is, then, given by the approximate expression
$P_{KN}\propto \pi r_e^2m^2c^5n_{ph}(\epsilon_{ph})\mid
ln\left(4e\epsilon/m^2c^4\right)- 11/6 \mid$ \cite{KN} where $r_e$
is the classical electron radius. The corresponding time scale of
the process is $t_{KN}\sim \epsilon /P_{KN}$.  Using the approximate
photon number density, $n_{ph}(\epsilon_{ph})\approx L/4\pi
R_{lc}^2c\epsilon_{ph}$, and taking into account the typical values
of luminosity in the high energy level, $\epsilon_{ph}\sim 10$GeV,
satisfy $L > 10^{35}$erg/s, it is straightforward to show that for
energies of our interest $\epsilon>1$PeV, the aforementioned
timescale exceeds the instability time scale by many orders of
magnitude. Thus Compton cooling is very slow compared to the wave
energy transfer time. The KN time scale goes up with the particle
energy implying that at higher energies, the inverse Compton
mechanism in the KN channel is too slow to impose any constraints on
the maximum attainable electron energy.

The next possible loss mechanism for relativistic particles, moving
in magnetic field, is the synchrotron emission with the
corresponding estimated power $P_{syn}\approx
2e^2\omega_B^2\gamma^2/3c$ \cite{Lightman}, where $\omega_B\equiv
eB/(mc)$ is the cyclotron frequency. According to the standard model
of pulsar magnetospheres, there is a significant  region over the
star's surface, where the electric field is nonzero (vacuum gap). It
is this electric field that, first, accelerates  electrons to
relativistic energies \cite{ruderman}. One can readily show that the
electrons leaving the gap with a $\gamma\sim 10^{6}$ will undergo
efficient synchrotron cooling at  a short timescale, $t_{syn}\sim
\gamma mc^2/P_{syn}= 10^{-21}$s. Thus, immediately after leaving the
gap, the particles radiate their transverse momentum, and very soon
transit to the ground Landau state. After that, zipping only along
the field lines, the electrons will reach the light cylinder zone in
due course of time. It is precisely the region, where the Langmuir
waves, always propagating along the field lines, are excited.
Therefore the wave interaction with particles will not cause pitch
angle scattering, efficiently suppressing the synchrotron mechanism.
Quasi linear diffusion, another possible source for imparting a
pitch angle, \cite{lmm}, also  does not operate because the required
condition for diffusion $\omega_B>\omega_{1,2}$, is violated for
extremely energetic plasmas (for plasmas with energy density
exceeding that of magnetic field). the synchrotron mechanism is not
expected to impose any constraints on the maximum attainable
energies.

How about the curvature radiation emitted by particles moving along
the curved magnetic field lines? To explain force free regime of
particles leaving a pulsar's magnetosphere, we have tried to
reconstruct the structure of magnetic field close to the light
cylinder surface \cite{osm09}. We showed that shown that the
curvature driven current  imparts a  toroidal component to the
magnetic field. As a result, the field lines lag behind the
rotation, gradually erasing the instability. It has been shown that
the corresponding timescale is approximately given by
\begin{equation}
\label{CDI}  \tau_{_{CD}}\approx
\left(\frac{2}{3}\frac{\gamma_b}{\omega^2}\frac{|k_{_{\perp}}|}{|k_{_{\parallel}}|}
\frac{c}{u_b}\right)^{1/2} \left[J_0\left(\frac{k_{_{\parallel}}
u_b}{4\Omega}\right)J_0\left(\frac{k_{_{\perp}}
c}{\Omega}\right)\right]^{-1},
\end{equation}
where $u_b\equiv\gamma_b c^2/(\omega_BR_B)$ is the curvature drift
velocity, $R_B\approx R_{lc}$ is the curvature radius of a magnetic
field line and $k_{_{\parallel}}$ and $k_{_{\perp}}$ are the
components of the wave vector of the induced curvature drift mode
respectively along the drift and perpendicular to it. One can see
that for a wide range of parameters corresponding to the drift mode
with small inclination angles $\alpha$ ( the angle of the wave
vector with respect to the drift direction), the instability
timescale is less than the kinematic timescale $P$. In particular,
for $\alpha\approx 0.2$ and $R_{lc}k_{_{\perp}}\sim 0.1$,
$\tau_{_{CD}}/\tau_{kin}\ll 1$. Therefore, the drift instability is
so efficient that the magnetic field lines very rapidly reconstruct
resulting in  a configuration, that enables particles to follow
straight line trajectories. The curvature radiation is minimized and
does not quite interfere
 with our energy transfer mechanism.

\begin{figure}
  \centering {\includegraphics[width=7cm]{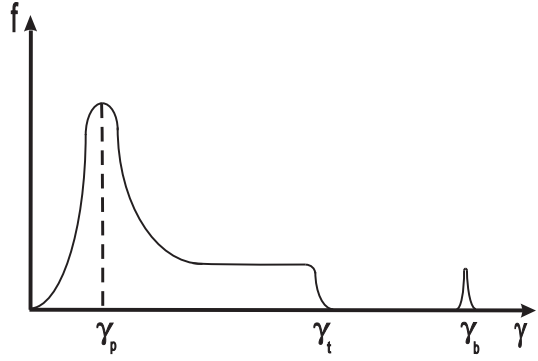}}
  \caption{The distribution function versus the Lorentz factor.
  It is evident that the function consists of two major parts: relatively
  narrower region with higher Lorentz factors, characterizes the primary
  Goldreich-Julian beam electrons and the wider part desc ribes the
  distribution of electrons resulting from the cascade processes of
  pair creation.}\label{fig1}
\end{figure}


{\bf Methods}

The LLCD was developed using the standard model of the pulsar
magnetosphere, where the particle distribution can be,
schematically, represented as shown in Fig. \ref{fig1}. The narrower
region on the figure describes the primary beam electrons with
higher Lorentz factors ($\gamma_b$).  The relatively wider sector
with Lorentz factors in the interval $[1; \gamma_t]$
($\gamma_t<\gamma_b$) characterizes the secondary particles produced
by means of pair creation. Such a shape of the distribution function
makes the development of two stream instabilities feasible in the
magnetospheres of pulsars \cite{gedalin}. This feature of the
electron distributions, along with interesting orbit effects, was
exploited in \cite{screp} to demonstrate the the existence of a
strong two-stream like instability that is parametrically driven to
excite high levels of Langmuir waves .

In the framework of $1+1$ formalism \cite{membran}, the basic system
of the Euler equation and continuity equations (for each species)
and the Poisson equation, is adequate in describing the
centrifugally-excited (rotationally driven) Langmuir waves. The
Fourier transformed linear system consists of
\begin{equation}
\label{eul3} \frac{\partial p_{_{\beta}}}{\partial
t}+ik\upsilon_{_{\beta0}}p_{_{\beta}}=
\upsilon_{_{\beta0}}\Omega^2r_{_{\beta}}p_{_{\beta}}+\frac{e_{_{\beta}}}{m}E,
\end{equation}
\begin{equation}
\label{cont1} \frac{\partial n_{_{\beta}}}{\partial
t}+ik\upsilon_{_{{\beta}0}}n_{_{\beta}}, +
ikn_{_{\beta0}}\upsilon_{_{\beta}}=0
\end{equation}
\begin{equation}
\label{pois1} ikE=4\pi\sum_{\beta}n_{_{\beta0}}e_{_{\beta}},
\end{equation}
where ${\beta}$ is the species index (electrons and positrons),
$p_{_{\beta}}$ and $\upsilon_{_{\beta}}$ are, respectively, the
first order dimensionless momentum ($p_{_{\beta}}\rightarrow
p_{_{\beta}}/m$) and the zeroth order velocity, $m$ is the electron
mass, $c$ is the speed of light, $e_{\beta}$ is the charge of the
corresponding particle, $r_{_{\beta}}$ is the radial coordinate of
the corresponding specie, $n_{_{\beta}}$ and $n_{_{\beta0}}$ are
respectively the perturbed and nonperturbed Fourier components of
the density, $k$ is the wave number and $E$ is the electrostatic
field. The first term in the righthand side of the Euler equation
comes from the so-called centrifugal force. The centrifugal force,
controlled by the  following "orbits" of initially relativistic
particles $r_{_{\beta}}(t) \approx
\left(V_{_{0\beta}}/\Omega\right)\sin\left(\Omega t +
\phi_{_{\beta}}\right)$, $\upsilon_{_{0\beta}}(t) \approx
V_{_{0\beta}}\cos\left(\Omega t + \phi_{_{\beta}}\right)$
\cite{screp}, where $\phi$ denotes the initial phase of a given
species, is time dependent. The resulting Langmuir mode equation is
periodic in time (has the generalized Mathieu/Hill form) and
exhibits parametric instability.

%

The time dependent centrifugal force that parametrically drives the
electrostatic waves is different for the two species - so are their
Lorentz factors.

Since ${\mu}_res$, the order of the Bessel function is rather high,
non-zero growth rates will pertain only if the argument $b$ is
comparable to ${\mu}_res$ \cite{abrsteg}.

\begin{figure}
  \centering {\includegraphics[width=7cm]{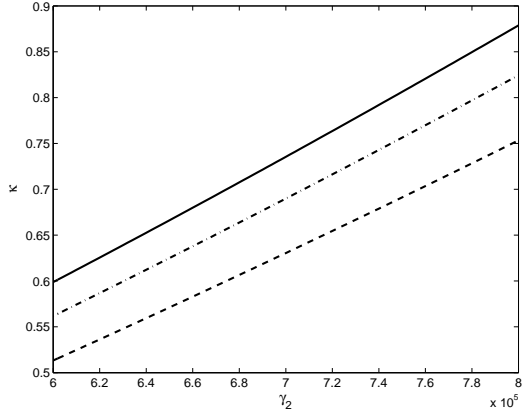}}
  \caption{Here we plot the dependence of $\kappa$ on $\gamma_2$ for three different values
  of $\gamma_1$: $\gamma_1 = 10^5$ (dashed-dotted line); $\gamma_1 = 1.5\times 10^5$ (dashed line) and
  $\gamma_1 = 2\times 10^5$ (solid line). As it is clear from the graph, for a quite wide range of
  Lorentz factors the instability timescale is less than the kinematic timescale,
  indicating high efficiency of the process.}\label{fig2}
\end{figure}
For the instability to be called "efficient", the inverse growth
time $\tau_{ins} = 1/\Gamma$ should not exceed the kinematic
timescale (also called the escape timescale), $\tau_{kin}\sim P$. In
Fig. \ref{fig2}, we plot  the ratio
$\kappa\equiv\tau_{ins}/\tau_{kin}$ versus $\gamma_2$ for different
values of $\gamma_1$: $\gamma_1 = 10^5$ (dashed-dotted line);
$\gamma_1 = 1.5\times 10^5$ (dashed line) and $\gamma_1 = 2\times
10^5$ (solid line). The results are derived  for typical young
millisecond pulsars with $P\sim 10^{-3}$s and $B_{lc}\approx
B_{\star}\times R_{\star}^3/R_{lc}^3$, where $B_{\star}\approx
10^{12}$G is the magnetic field close to the pulsar's surface and
$R_{lc}\equiv c/\Omega$ is the light cylinder radius. $\dot{P} \sim
10^{-12}$ss$^{-1}$. The condition $\tau_{ins}<\tau_{kin}$ is
satisfied for a broad range of Lorentz factors; the instability is,
indeed, efficient.

Thus we have reviewed, for the young millisecond pulsars, the first
step of LLCD: the existence of an efficient centrifugally excited
parametric instability that pumps the spin down energy of the
rotator into electrostatic Langmuir waves. Now we review how,
through landau damping, this energy will be conveyed to the fastest
of the particles. The Landau damping in pulsar magnetospheres has
been investigated in \cite{arons}, but the authors focus on Alfven
waves, but not on the electrostatic waves.


\section*{Acknowledgments}
The research of ZO, GM and NC was partially supported by the Shota
Rustaveli National Science Foundation grant (N31/49). The work of SM
was, in part, supported by USDOE Contract No.DE-- FG 03-96ER-54366.

{\bf Affiliations}

School of Physics, Free University of Tbilisi, 0183-Tbilisi,
Georgia\\
Zaza Osmanov\\
Institute for Fusion Studies, The University of Texas at Austin,
Austin, Texas\\
Swadesh Mahajan\\
Centre for Theoretical Astrophysics, ITP, Ilia State University,
0162-Tbilisi, Georgia\\
George Machabeli, Nino Chkheidze

{\bf Contributions}

ZO conceived the project; ZO, SM, GM and NC developed the analytical
model. ZO performed the numerical investigation of the problem, who
also wrote the manuscript with the assistance of all other
co-authors.

{\bf Competing financial interests}

The authors declare no competing financial interests.

{\bf Corresponding author}

Correspondence to: Zaza Osmanov


\begin{thebibliography}{999}

\bibitem{hesstev} Aharonian, F., et al., Energy spectrum of cosmic-ray
electrons at TeV Energies. {\it Phys. Rev. L.}, {\bf 101}, 261104,
(2008)
\bibitem{fermitev} Ackermann, M., et al. Fermi LAT observations of
cosmic-ray electrons from 7 GeV to 1 TeV. {\it Phys. Rev. D.}, {\bf
82}, 092004, (2010)
\bibitem{bell} Bell, A. R., The acceleration of cosmic rays in shock fronts.
{\it MNRAS}, {\bf 182}, 147-156, (1978)
\bibitem{blandford} Blandford, R. D., Netzer, H. \& Woltjer, L.
{\it Active Galactic Nuclei}, Springer-Verlag (1990)
\bibitem{screp}
Mahajan, S., Machabeli, G., Osmanov, Z. \& Chkheidze, N., Ultra high
energy electrons powered by pulsar rotation. {\it Nat. Sci. Rep.}
{\bf 3}, 1262, (2013)
\bibitem{bogoval2}
Bogovalov, S., Magnetocentrifugal acceleration of plasma in a
nonaxisymmetric magnetosphere. {\it A\&A}, {\bf 367}, 159-169,
(2001)
\bibitem{osm7} Osmanov, Z., Rogava, A.S. \& Bodo, G.,
On the efficiency of particle acceleration by rotating
magnetospheres in AGN. {\it A\&A}, {\bf 470}, 395-400, (2007)
\bibitem{youngPL} Fang, K., Kotera, K. \& Olinto, A. V.,
Newly Born Pulsars as Sources of Ultrahigh Energy Cosmic Rays. {\it
ApJ}, {\bf 750}, 1-16, (2012)
\bibitem{vkm}
Volokitin, A.S., Krasnoselskikh, V.V. \& Machabeli, G.Z., Waves in
the relativistic electron-positron plasma of a pulsar. {\it SvJPP},
{\bf 11}, 310-314, (1985)

\bibitem{carroll} Carroll, B.W. \& Ostlie, D.A. {\it An Introduction
to Modern Astrophysics}, Pearson International Edition (2006)


\bibitem{grg} Rogava, A., Dalakishvili, G. \&
Osmanov, Z., Centrifugally driven relativistic dynamics on curved
trajectories. {\it GReGr}, {\bf 35}, 1133-1152, (2003)
\bibitem{gj} Goldreich, P. \&
Julian, W. H., Pulsar electrodynamics. {\it ApJ}, {\bf 157},
869-880, (1969)
\bibitem{Lightman} Rybicki,  G.B. \& Lightman, A. P.,
{\it Radiative Processes in Astrophysics}. Wiley, New York, (1979)
\bibitem{KN} Blumenthal, G. R. \& Gould, R. J.,
Bremsstrahlung, synchrotron radiation, and compton scattering of
high-energy electrons traversing dilute gases. {\it Rev. Mod.
Phys.}, {\bf 42}, 237-271, (1970)
\bibitem{ruderman} Ruderman, M. A. \&
Sutherland, P. G., Theory of pulsars - polar caps, sparks, and
coherent microwave radiation. {\it ApJ}, {\bf 196}, 51-72, (1975)
\bibitem{lmm} Lominadze J.G., Machabeli G.Z. \& Mikhailovsky A.B., 1979,
Relativistic electron-positron plasma quasi-linear relaxation at the
presence of magnetic bremsstrahlung. {\it J. Phys. Colloq.}, {\bf
40}, No. C-7, 713, (1979)
\bibitem{osm09} Osmanov, Z., Shapakidze, D. \& Machabeli G.,
Dynamical feedback of the curvature drift instability on its
saturation process. {\it A\&A}, {\bf 503}, 19-24, (2009)
\bibitem{gedalin} Gedalin, M., Gruman, E. \& Melrose D. B.,
New mechanism of pulsar radio emission. {\it Phys. Rev. L.}, {\bf
88}, 121101, (2002)
\bibitem{membran} Thorne, K., Price, R. \& MacDonald, D.A. {\it Black Holes: The
Membrane Paradigm}, Yale University Press, New Haven (1986)
\bibitem{abrsteg} Abramowitz, M. \&
Stegun, I. A., {\it Handbook of Mathematical Functions}, edited by
Abramowitz, M. \& Stegun, I. A., Natl. Bur. Stand. Appl. Math. Ser.
No. 55 (U.S. GPO, Washington, D.C., 1965)
\bibitem{arons} Arons, J. \& Scharlemann, E. T.,
Pair formation above pulsar polar caps - Structure of the low
altitude acceleration zone. {\it ApJ}, {\bf 231}, 854-879, (1979)





\end{thebibliography}
\end{document}